\documentclass[epj]{svjour}

\usepackage{graphicx}
\usepackage{dcolumn}
\usepackage{bm}
\usepackage{graphics}
\usepackage{latexsym}
\usepackage{graphics}
\usepackage{dcolumn}
\usepackage{epsfig}
\usepackage{amssymb} 
\usepackage{amsmath} 
\usepackage{rotate} 
\usepackage{color} 
\usepackage{graphicx}
\usepackage{amssymb}
\newcommand{\gammap}{\dot{\gamma}}
\newcommand{\gammapeff}{\gammap_{\hbox{\rm\scriptsize true}}}
\newcommand{\heq}{h_{\hbox{\rm\scriptsize eq}}}
\begin{document}
\title{A spatio-temporal study of rheo-oscillations\\
in a sheared lamellar phase using ultrasound}
\author{S\'ebastien Manneville \and Jean-Baptiste Salmon \and Annie Colin}
\institute{Centre de Recherche Paul Pascal, Avenue Schweitzer, 33600 Pessac, FRANCE}


\date{\today}

\abstract{We present an experimental study of the flow dynamics of a lamellar phase sheared in the Couette geometry.
High-frequency ultrasonic pulses at 36~MHz are used to measure time-resolved velocity profiles.
Oscillations of the viscosity occur in the vicinity of a shear-induced transition
between a high-viscosity disordered fluid and a low-viscosity ordered fluid. The phase
coexistence shows up
as shear bands on the velocity profiles.
We show that the dynamics of the rheological data result from two different processes:
(i) fluctuations of slip velocities at the two walls and (ii) flow dynamics in the bulk of the lamellar phase. The bulk
dynamics are shown to be related to the displacement of the interface between the two differently sheared
regions in the gap of the Couette cell.
Two different dynamical regimes are investigated under applied shear stress: one of small amplitude oscillations
of the viscosity ($\delta\eta/\eta\simeq 3$~\%) and one of large oscillations ($\delta\eta/\eta\simeq 25$~\%).
A phenomenological model is proposed that may account
for the observed spatio-temporal dynamics.}


\authorrunning{S. Manneville \textit{et al.}}

\titlerunning{Spatio-temporal study of rheo-oscillations using ultrasound}

\maketitle

\section{Introduction}

When submitted to a shear flow, complex fluids exhibit unusual behaviour: steady shear modifies their structure 
and induces new structures or textures that do not exist at rest \cite{Larson:1999}. In wormlike micelles for instance,
 shear may induce a nematic phase \cite{Berret:1994b,Schmitt:1994}, whereas in membrane phases, shear creates new textural organizations
such as multilamellar vesicles \cite{Diat:1993a,Diat:1993b}.
 Those shear-induced effects may be summarized in a {\it shear diagram} that specifies the nature of the stationary 
states (texture or phase) observed for different shear rates (or shear stresses).
These states are separated by dynamical transitions that
correspond to a jump between two branches of steady states \cite{Roux:1993}.
In some cases, the fact that the complex fluid is sheared,
and thus maintained out of equilibrium, leads to richer behaviours than simple stationary states. 
Indeed, transitions to oscillating states and to chaotic-like behaviours have been reported recently in various systems
\cite{Hu:1998a,Wheeler:1998,Bandyopadhyay:2000,Wunenburger:2001,Salmon:2002,Hilliou:2002,Lootens:2003,Courbin:2003}.
These transitions involve rich dynamics of the fluid microstructure, which are often coined ``rheochaos" \cite{Cates:2002}, and
differ from classical instabilities, such as the Taylor-Couette instability \cite{Tritton:1988}
or elastic instabilities \cite{Shaqfeh:1996}.

In this work, we study a particular complex fluid exhibiting such {\it structural dynamics}:
a lamellar phase composed of sodium dodecyl sulfate (SDS), octanol, and brine.
The shear diagram of this lyotropic system depends upon the temperature. For 
small applied shear rates $\gammap<\gammap_1$, the membranes are rolled into multilamellar
vesicles called ``onions", that form a {\it disordered} close-compact
monodisperse assembly \cite{Sierro:1997}. 
The characteristic size of the onions is a few microns. 
For large applied shear rates $\gammap > \gammap_2 > \gammap_1$ and at temperatures $T\simeq 30~^\circ$C,
onions are {\it hexagonally ordered} into layers that lie in the velocity-vorticity plane normally to the shear
gradient and slide on one another. 
This structural transition, called the ``layering" transition, is shear-thinning and occurs between
$\gammap_1$ and $\gammap_2$ at a given critical shear stress $\sigma_c$ \cite{Sierro:1997}.
The rheological flow curve (measured shear stress $\sigma$ vs. applied shear rate $\gammap$)
thus shows very strong shear-thinning around $\sigma_c$ that leads to
a quasi-horizontal plateau at $\sigma_c$. Such a rheological behaviour seems
generic in complex fluids as soon as a shear-induced transition is involved
\cite{Rehage:1991,Eiser:2000,Ramos:2001a,Berret:1997} and has been much studied in the framework
of shear banding theories \cite{Spenley:1993,Olmsted:1997}.

Under {\it imposed shear stress}, the system follows the same rheological flow curve. Stationary states are reached
at low and high shear stresses below and above $\sigma_c$.
However, in a narrow region in the vicinity of the critical shear stress, complex temporal
rheological behaviours are recorded and no stationary state is ever reached
\cite{Wunenburger:2001,Salmon:2002}.
These temporal behaviours correspond to structural changes of one part or of the entire sample between
a disordered and an ordered assembly of multilamellar vesicles.

In order to obtain good reproducibility in the flow curve $\sigma$ vs. $\gammap$, careful procedures must be followed:
the applied shear stress must be maintained during time intervals $\delta\tau$ longer than 1000~s and the increment between two 
different imposed shear stresses $\delta \sigma$ must be smaller than 1~Pa. However,
the obtained state still depends upon the value of $\delta\sigma$ and $\delta\tau$.
Using a rather fast procedure, we have evidenced in the plane $\sigma$ vs. $\gammap$ a succession of stationary states,
noisy regions, regions of small amplitude oscillations
($\delta\gammap/\gammap=\delta\eta/\eta\simeq 3$~\% \cite{Remark:fluctu}) and of large metastable oscillations
($\delta\gammap/\gammap\simeq 25$~\%) as the shear stress is increased.
A more ``quasistatic" procedure reveals chaotic-like
aperiodic signals and the large oscillations are no longer observed, which confirms their metastability.
Using dynamical system theory, we have shown that the chaotic-like oscillations of the
viscosity may not be simply described by a three-dimensional dynamical system, 
probably because spatio-temporal effects play a crucial role \cite{Salmon:2002}.

In order to go further into the understanding of such behaviours, local velocity
profiles are required to probe the flow inside the gap of a Couette cell (concentric cylinders).
Using dynamic light scattering (DLS) experiments in the heterodyne geometry, we have shown that a simple shear banding scenario holds for
temporally averaged velocity profiles along
the layering transition under applied shear rate \cite{Salmon:2003d}:
the low-viscosity ordered phase nucleates
at the rotor and progressively fills the entire gap
of the Couette cell as the applied shear rate 
is increased from $\gammap_1$ to $\gammap_2$. In this range of applied shear rates, the two phases coexist and the flow is {\it inhomogeneous}. 
The temporally averaged position of the interface between the two phases lies at a fixed shear stress $\sigma^\star$. 
However, noticeable fluctuations of the band position have been reported \cite{Salmon:2003e}.
The poor temporal resolution of our DLS setup has prevented us to further study the dynamics under applied shear 
stress.
Indeed, about three minutes are required to measure a complete profile with the DLS technique \cite{Salmon:2003b}.
Note that this resolution is poor but comparable to that of Nuclear Magnetic Resonance (NMR) velocimetry
with the same spatial resolution of 50~$\mu$m \cite{Callaghan:1991,Hanlon:1998}.

In this work, we present an experimental study of the layering transition under {\it imposed shear stress}
using a recently developed experimental setup. High-frequency ultrasonic pulses at 36~MHz are used to measure velocity profiles. 
Our technique is based on time-domain cross-correlation of high-frequency ultrasonic signals backscattered 
by the moving fluid. Post-processing of acoustic data allows us to record a velocity profile 
in 0.02--2~s with a spatial resolution of 40~$\mu$m \cite{Manneville:2003pp_a}.
Such a temporal resolution allows us to follow the dynamics of velocity profiles during the viscosity
oscillations and to better understand the mechanisms at play during these oscillations.
The article is organized as follows. In section~\ref{s.exp}, we briefly describe the system under study
as well as the experimental setup for measuring time-resolved velocity profiles in a Couette cell. 
Section~\ref{s.small} describes the experiments performed in the region of small amplitude oscillations.
Section~\ref{s.large} deals with the large metastable oscillations. Finally, Sec.~\ref{s.discuss}
is devoted to a discussion of our experimental results and a phenomenological model is proposed
that may account for the observed spatio-temporal dynamics.


\section{Experimental section }
\label{s.exp}

\subsection{System under study}

The complex fluid investigated in this work is made of sodium dodecyl sulfate (SDS), octanol, and brine. At the concentrations
 considered here (6.5~\%~wt. SDS, 7.8~\%~wt. octanol, and 85.7~\%~wt. brine at 20~g.L$^{-1}$), a lamellar phase is observed
at equilibrium. The smectic period is 15~nm and the bilayer thickness
is about 2~nm \cite{Herve:1993}. For the given range of concentrations, the lamellar phase is stabilized by undulating
interactions. This system is very sensitive to temperature: for $T\gtrsim 35~^\circ$C,
a sponge--lamellar phase mixture appears.

Moreover, this lamellar phase is transparent to ultrasonic waves and does not significantly scatter 36~MHz pulses,
Thus, in order to measure the velocity profiles, the sample has to be seeded with latex spheres of diameter of a few microns (see Sec.~\ref{s.setup} below).
These spheres are obtained from polymerization of an emulsion of divinylbenzene (DVB) and benzoyl peroxide
(respectively 39.6~\%~wt. and 0.4~\%~wt.) in water stabilized by SDS (1~\%~wt.) at a temperature of 
80~$^\circ$C \cite{Echevarria:1998}. Benzoyl peroxide was added to the DVB to initiate the reaction.
This emulsion is then concentrated by centrifugation, diluted in the lamellar phase, and again concentrated. Finally,
1~\% wt. of the obtained paste is suspended in the lamellar phase.
Such a washing process allows us to avoid any modification of the lamellar composition.
We checked that the addition of latex spheres does not significantly affect the rheological behaviour and the
layering transition of our system.

\subsection{Experimental setup}
\label{s.setup}

\begin{figure}[htbp]
\begin{center}
\scalebox{1}{\includegraphics{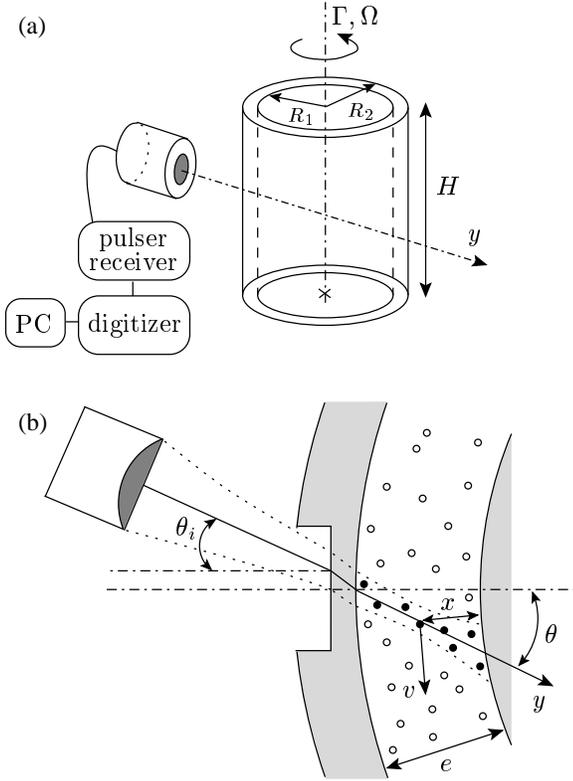}}
\end{center}
\caption{\label{f.setup}Experimental setup. (a) General sketch of the experiment. (b) Enlarged view of the gap of the Couette cell as seen from above.
Dotted lines represent the ultrasonic beam and black dots ($\bullet$) indicate the latex spheres that contribute to the backscattered signal.}
\end{figure}

To study the effect of shear flow on this lamellar phase, we used the experimental device sketched in Fig.~\ref{f.setup}(a)
and described at length in Ref.~\cite{Manneville:2003pp_a}. A rheometer (TA Instruments AR1000N)
allows us to perform rheological measurements in a Couette cell.
The Couette cell is made of Plexiglas and has the following characteristics:
inner radius $R_1=24$~mm, outer radius $R_2=25.07$~mm, gap $e=R_2-R_1=1.07$~mm, and height $H=30$~mm.
The walls of both cylinders are smooth. The rheometer imposes a constant torque $\Gamma$  on the axis of the inner cylinder,
which induces a constant stress $\sigma$ in the fluid, and measures
the rotation speed $\Omega$ of the rotor (rotating inner cylinder) .
A computer-controlled feedback loop on the applied torque $\Gamma$
can also be used to apply a constant shear rate $\gammap$ without any signifiant temporal
fluctuations. The relations between the ``engineering" ({\it i.e.} global) quantities
($\sigma,\gammap$) given by the rheometer and ($\Gamma$,$\Omega$) read
\begin{eqnarray}
\sigma&=&\frac{R_1^2+R_2^2}{4\pi HR_1^2R_2^2}\,\Gamma\,,\label{e.sigmarheo}\\
\gammap&=&\frac{R_1^2+R_2^2}{R_2^2-R_1^2}\,\Omega\, .
\label{e.gammarheo}
\end{eqnarray}
Such definitions ensure that ($\sigma,\gammap$) correspond to the spatially
averaged values of the local stress and shear rate in the case of a Newtonian fluid.
The whole cell is surrounded by water whose temperature is kept constant to within $\pm 0.1~^\circ$C.

The thickness of the stator (fixed outer cylinder) is 2~mm 
everywhere except for a small rectangular window where the minimal thickness is 0.5~mm in order to avoid additional attenuation
of the ultrasonic pulses due to Plexiglas.
A piezo-polymer transducer of central frequency $f=36$~MHz is immersed in the water in front of the stator. It generates ultrasonic pulses
focused in the gap of the cell with a given incidence angle $\theta_i\simeq 10^\circ$ relative to the normal to the window. 
The ultrasonic pulses travel through Plexiglas and enter the gap with an angle $\theta$ that is given by the law of refraction (see Fig.~\ref{f.setup}(b)).
The transducer is controlled by a pulser-receiver unit (Panametrics 5900PR).
The pulse repetition frequency is tunable from 0 to 20~kHz. Typically, pulses are separated by a few milliseconds.
Once inside the fluid, ultrasonic pulses get scattered by the latex spheres. 
Backscattered signals are sampled at 500~MHz, stored on a high-speed PCI digitizer (Acqiris DP235),
and transferred to the host computer for post-processing.
Typical recorded backscattered signals are 1000 point long, which corresponds to a transit time of 2~$\mu$s.

Under the assumption of single scattering, the signal received at time $t$ can be interpreted as interferences coming from
scattering particles located at position $y=c_0 t/2$, where $c_0$ is the sound speed
in the fluid and $y$ the distance from the transducer along the ultrasonic beam. 
When the fluid is submitted to a shear flow, the backscattered signals change as the scattering particles move along.
Two successive pulses separated by $\Delta T$ lead to two backscattered signals that are shifted in time.
The time-shift $\delta t$ between two echoes received at time $t$ corresponds to the displacement $\delta y = c_0 \delta t/2$
along the $y$-axis of scattering particles initially located at position $y=c_0 t/2$. The velocity $v_y$, projected along the $y$-axis, of the scatttering
particles located at $y$ is then simply given by $v_y=\delta y/\Delta T$.
A cross-correlation algorithm is used to estimate the time-shift $\delta t$ as a function of $t$.
The sound speed $c_0$ is measured independently, which yields the projection $v_y(y)$ in the entire gap.

Finally, a calibration procedure using a Newtonian fluid is required to
find precisely the angle $\theta$, so that velocity profiles $v(x)$
of the tangential velocity $v$ as a function of the radial distance $x$ to the rotor can be computed from $v_y(y)$ under the assumtion that the flow remains two-dimensional and axisymmetric.
The reader is referred to Ref.~\cite{Manneville:2003pp_a} for more technical details.
Typical velocity profiles are obtained by averaging over 50 series of twenty consecutive pulses.
The spatial resolution of our experimental setup is of the order of 40~$\mu$m and the temporal resolution ranges between 0.02 and 2~s per profile.


\section{Small amplitude oscillations}
\label{s.small}

\subsection{Experimental procedure and rheological measurements}

To understand how the system flows when oscillations of the viscosity occur, we record the ``engineering" ({\it i.e.} global) 
rheological data simultaneously to the velocity profiles. In order to obtain reproducible experiments, careful rheological protocols must be followed
as discussed in Ref.~\cite{Salmon:2002}. At the temperature under study ($T=32~^\circ$C),
we apply a first step of imposed shear stress at $\sigma=15$~Pa during 6300~s. This step of applied shear stress
allows us to start the experiment with a well-defined stationary state of disordered onions. 
We then apply increasing shear stresses for at least 2800~s per step. The stress increment between two steps is $\delta\sigma=1$~Pa.

\begin{figure}[htbp]
\begin{center}
\scalebox{1}{\includegraphics{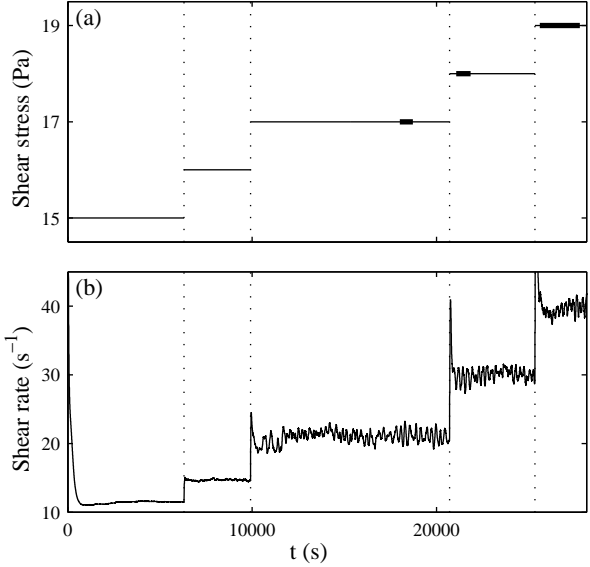}}
\end{center}
\caption{\label{f.rheol2}(a) Imposed engineering shear stress $\sigma(t)$. Thick lines correspond to the time intervals
when velocity profiles were recorded. (b) Measured engineering shear rate $\gammap(t)$.
The dotted lines correspond to the end of the various steps.
The temperature is $T=32~^\circ$C.}
\end{figure}

Engineering rheological measurements are displayed in Fig.~\ref{f.rheol2}. In the imposed shear stress mode, the temporal fluctuations
of the shear stress are completely negligible.
For the first two steps, the system reaches a stationary state ($\delta\gammap/\gammap\lesssim 0.7$~\%). However, for imposed shear stresses greater than 16~Pa,
noticeable {\it fluctuations} of the measured shear rate (and thus of the global viscosity of the sample) are recorded: $\delta\gammap/\gammap\simeq 3$~\%.
The characteristic times involved in such temporal behaviours are
in the range 50--500~s as reported in previous experiments \cite{Wunenburger:2001,Salmon:2002}. Moreover, using static light scattering, we checked that
the onset of these temporal fluctuations corresponds to the apparition of the ordered texture in the sample (see Refs.~\cite{Wunenburger:2001,Salmon:2002}
for more details). Finally, note that all the experiments discussed here and in previous papers 
are conducted far from any hydrodynamic or elastic
instability (such instabilities would, in any case, lead to much shorter time scales), so that the observed temporal behaviours can 
unambiguously be attributed to the presence of a shear-induced structural instability in our lamellar phase.

\subsection{Time-averaged velocity profiles}
\label{s.small.average}

\begin{figure}[htbp]
\begin{center}
\scalebox{1}{\includegraphics{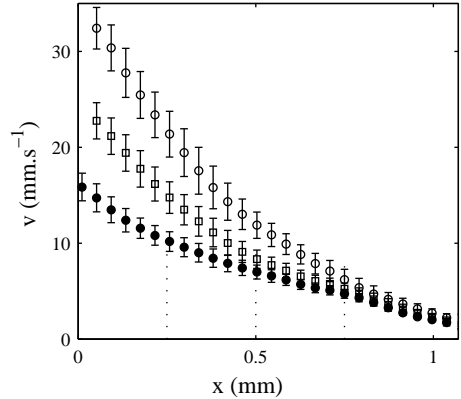}}
\end{center}
\caption{\label{f.moy2}Time-averaged velocity profiles for $\sigma=17$ ($\bullet$), 18 ($\square$), and
19~Pa ($\circ$) recorded simultaneously to the rheological data displayed in Fig.~\ref{f.rheol2}.
Error bars correspond to the standard deviation of the local velocities and mainly
account for temporal fluctuations of the velocity field (see text). The dotted lines
show the average position of the interface between the two shear bands. The mean velocity of the rotor 
is $v_0=20.7$, 28.8, and 38.7~mm.s$^{-1}$ respectively. Wall slip occurs both at the rotor and at the stator (see text).}
\end{figure}

In Fig.~\ref{f.moy2}, we present a time average of all the tangential velocity profiles $v(x)$ measured for the different applied shear stresses.
$x$ denotes the radial distance to the rotor ($x=0$ at the rotor and $x=e$ at the stator).
For each applied shear stress, the time intervals when velocity profiles were measured are indicated by thick lines in Fig.~\ref{f.rheol2}(a).
Typically one hundred profiles (recorded every 5~s  in about 1~s per profile) have been averaged to obtain the data of Fig.~\ref{f.moy2}.
The error bars correspond to the standard deviation of
the velocity measurements. They are much larger than the uncertainty on individual
measurements and are due to strong temporal fluctuations of the velocity field.
These profiles clearly reveal an inhomogeneous flow. Two differently sheared bands with different viscosities coexist in the gap of the Couette cell.
The dotted lines in Fig.~\ref{f.moy2} indicate the average position of the interface between the two regions of different viscosities.
When the stress is increased, the highly sheared band expands across the gap.
This picture is in agreement with the classical {\it shear banding} scenario \cite{Spenley:1993,Olmsted:1997} and with previous 
measurements obtained on the same lamellar phase (without seeding particles) using DLS \cite{Salmon:2003d}.

Moreover, {\it significant wall slip} can be detected on these measurements: the velocity at the stator 
does not vanish at $x=e$ and the velocity does not reach the rotor velocity $v_0=R_1\Omega$ at $x=0$. Let us define the slip velocity at 
the rotor $v_{s1}=v_0-v_1$ as the difference between the rotor velocity $v_0$ and the velocity $v_1$ of the fluid close to the rotor,
and the slip velocity at the stator $v_{s2}=v_2$ as the velocity $v_2$ of the fluid near the stator.
Experimentally, $v_1$ and $v_2$ are estimated by linear fits of the velocity profiles over 100~$\mu$m close to the walls.
At the rotor, one gets $v_{s1}=4.5$, 4.1, and 3.8~mm.s$^{-1}$ for engineering shear stresses 
$\sigma=17$, 18, and 19~Pa respectively, which correspond to local shear stresses at the rotor
$\sigma_1=17.7$, 18.8, and 19.8~Pa respectively. At the stator, $v_{s2}=1.4$, 1.7, and 1.9~mm.s$^{-1}$ for local shear stresses
at the stator $\sigma_2=16.3$, 17.2, and 18.2~Pa respectively. The local shear stresses have been calculated using the following relationship:
\begin{equation}
\sigma_i=\frac{2R_j^2}{R_1^2+R_2^2}\,\,\sigma\,,
\label{e.sigmawall}
\end{equation}
where $j=2$ (resp. $j=1$) when $i=1$ (resp. $i=2$).
Note that it is important not to confuse the engineering value $\sigma$ with the local values $\sigma_1$, $\sigma_2$, and more
generally $\sigma(x)=\sigma_1 R_1^2/(R_1+x)^2$.
In the coexistence domain of the layering transition, we thus note a large difference between slip velocities at the rotor and at the stator:
wall slip is much stronger at the rotor than at the stator.
These results are in qualitative agreement with previous DLS experiments showing the existence of lubricating layers of thickness of about
50--250~nm with larger slip velocities at the rotor than at the stator \cite{Salmon:2003d}.

In the following section, we focus on the individual profiles measured at
$\sigma=19$~Pa and try to capture the details of the flow dynamics.

\subsection{Study of the step at $\sigma=19$~Pa}

In this part of the article, we present a  complete study of the last step at an imposed shear stress of 19~Pa.
We chose to study this particular step because of the large number of velocity profiles recorded during this step
(415 profiles separated by 5~s). Note, however, that the behaviour observed at 19~Pa is qualitatively the same for the two other steps
where similar complex dynamics of $\gammap(t)$ are recorded ($\sigma=17$ and 18~Pa).

\subsubsection{Local shear rate and interface position}
\label{s.sptp}

\begin{figure}[htbp]
\begin{center}
\scalebox{1}{\includegraphics{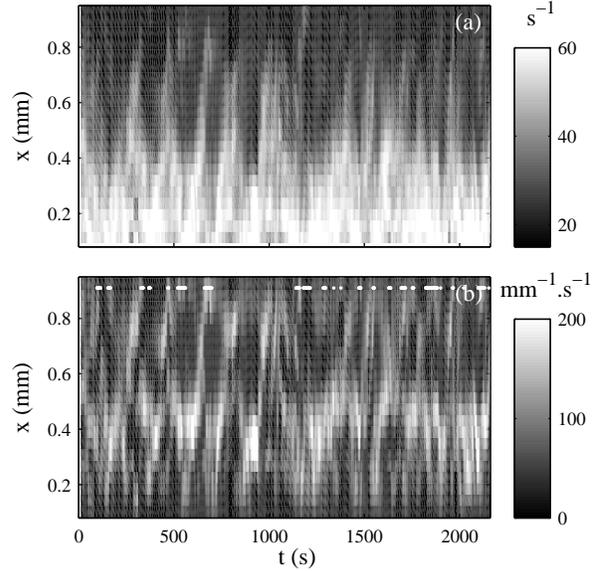}}
\end{center}
\caption{\label{f.sptp2}(a) Local shear rate $\gammap(x,t)$.
(b) Spatial derivative of the local shear rate : $\partial\gammap(x,t)/\partial x$. 
Moving averages on both space and time were used to filter out noise due to differentiation of the experimental data.
White lines and dots indicate time intervals when two (or more) interfaces are present (see text).}
\end{figure}

Figure~\ref{f.sptp2}(a) presents a spatio-temporal diagram of the shear rate profiles $\gammap(x,t)$ \cite{Remark:moviesUSV}.
The local shear rate is given by
\begin{equation}
\gammap(x,t)=-(R_1+x)\,\frac{\partial}{\partial x}\left( \frac{v(x,t)}{R_1+x}\right)\,.
\end{equation}
Such a spatial derivative of the velocity field is very sensitive to experimental uncertainties
and we used a moving average on a space window of three consecutive points and on a temporal window of three consecutive profiles
to obtain the data shown in Fig.~\ref{f.sptp2}(a).
The abscissae in Fig.~\ref{f.sptp2}(a) correspond to time $t$ whereas the ordinates correspond to positions $x$ in the gap.
The value of the shear rate at a given time and a given position in the gap is coded using gray levels, darker values corresponding to smaller shear rates. 
Figure~\ref{f.sptp2}(a) clearly reveals that the flow is both {\it inhomogeneous}
and {\it non-stationary}. A highly sheared band
is located near the rotor ($x\simeq 0$--0.4~mm).
The interface between the two
shear bands is rather sharp and its position seems
to fluctuate strongly in time.

In order to present the evolution of the interface position in more details, we calculate the derivative of the shear rate profile as a function of space
 $\partial\gammap(x,t)/\partial x$. 
In the case of a two-band velocity profile, one expects such a derivative of the local shear rate to be close to zero in the bands and
to present a maximum in the region of the interface.
Again, $\partial\gammap(x,t)/\partial x$ had to be filtered by using a spatial moving average over four consecutive points
in order to reduce noise due to differentiation.
The resulting data shown in Fig.~\ref{f.sptp2}(b) reveal that
the position of the interface between the two bands oscillates inside the gap.
We also note that, in some particular cases indicated by white lines and dots in Fig.~\ref{f.sptp2}(b),
two (or more) maxima are found at a given time. 
These two maxima correspond to two interfaces and thus to the presence of three bands of different shear rates.
Such profiles with three (or more) shear bands have a short lifetime and seem highly unstable:
the interface near the stator either moves rapidly towards the interface located near the rotor or disappears at the stator (see below).

\subsubsection{Individual two-banded and multi-banded velocity profiles}

\begin{figure}[htbp]
\begin{center}
\scalebox{1}{\includegraphics{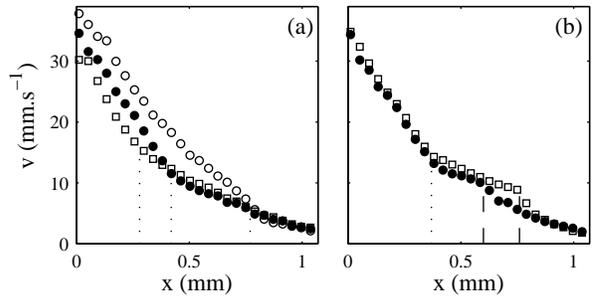}}
\end{center}
\caption{\label{f.prof}Individual velocity profiles recorded at $\sigma=19$~Pa. (a) $t=617$ ($\bullet$), 863 ($\square$), and 1000~s ($\circ$). 
The dotted lines show the position of the interface between the two shear bands.
(b) $t=525$ ($\square$) and 536~s ($\bullet$). The dotted line marks the position of
the main stable interface whereas dashed lines help to locate 
an unstable secondary interface near the stator. The experimental
uncertainty is of the order of the marker size.}
\end{figure}

In order to better illustrate the two points discussed above
(fluctuations of the interface position and presence of more than two shear bands in the Couette cell),
we isolate different individual velocity profiles in Fig.~\ref{f.prof}.
Figure~\ref{f.prof}(a) unambiguously shows that the interface between the two shear bands fluctuates: it is located at $x=0.42$~mm at $t=617$~s,
moves to $x=0.28$~mm at $t=863$~s, and goes as far as $x=0.77$~mm at $t=1000$~s. These profiles present a single interface
whereas three interfaces are clearly seen on the velocity profiles shown in Fig.~\ref{f.prof}(b).
At times $t=520$--540~s, a very narrow, highly sheared band is nucleated near the stator
and moves rapidly towards the rotor where it coalesces with the initial, more stable high-shear band located at $x\simeq 0.4$~mm.

From the data of Fig.~\ref{f.prof}(b), we may also roughly estimate the velocity of the secondary interface to be about
0.1~mm$/10$~s$\,\simeq 0.01$~mm.s$^{-1}$. On the other hand, the main interface travels about 1~mm in typically 100~s (see Fig.~\ref{f.sptp2}(b)),
which corresponds to an interface velocity of the same order of magnitude. Thus, if one assumes that the propagation
speed is characteristic of the interface \cite{Radulescu:1999}, we may infer that the various bands are of the same nature,
although secondary bands seem unstable. Note that this picture assumes that the flow remains two-dimensional and
axisymmetric so that the ultrasonic measurements can be interpreted in terms of the tangential velocity $v(x,t)$. However,
we cannot exclude the possibility of three-dimen\-sio\-nal events that could lead
to transient bumps in the velocity profiles such as those of Fig.~\ref{f.prof}(b).
Simultaneous imaging of two components of the velocity field is required to draw definite conclusions on that
specific issue and is left for future work.

\subsubsection{Wall slip and band dynamics}
\label{s.slip}

As shown above in Sec.~\ref{s.small.average}, wall slip
occurs both at the rotor and at the stator and the slip velocities $v_{s1}$ and $v_{s2}$ may well vary strongly in time. Thus,
in order to analyze further the flow dynamics, we first
need to measure quantitatively the slip velocities as a function of time, as well as the shear rate
in the bulk of the lamellar phase.

\begin{figure}[htbp]
\begin{center}
\scalebox{1}{\includegraphics{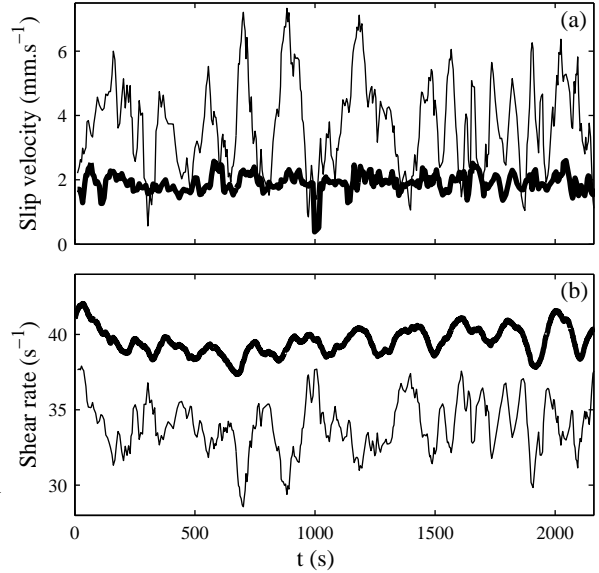}}
\end{center}
\caption{\label{f.gliss2}(a) Slip velocities at the rotor $v_{s1}(t)$ (thin line) and at the stator $v_{s2}(t)$ (thick line).
(b) Engineering shear rate $\gammap(t)$ (thick line) and true shear rate $\gammapeff(t)$ (thin line).
These data were filtered by using a moving average on a temporal window of three consecutive profiles.}
\end{figure}

Figure~\ref{f.gliss2}(a) presents the temporal evolution of $v_{s1}$ and $v_{s2}$. In order to avoid any side effects,
$v_{s1}(t)$ and $v_{s2}(t)$ are
obtained from linear fits of each velocity profile for $0 \leq x \leq 100~\mu$m and $e-150~\mu$m$\,\leq x \leq e-50~\mu$m respectively. 
As noticed above on the average profiles, sliding is stronger at the rotor than at the stator. Like $v(x,t)$, the two slip velocities
present important temporal fluctuations with characteristic times in the range 50--500~s. Slip dynamics in the two lubricating layers
are also strongly {\it dissymmetric} since the slip velocity at the rotor fluctuates by $\delta v_{s1}/v_{s1}\simeq 37$~\%, whereas 
$\delta v_{s2}/v_{s2}\simeq 15$~\% only. 

From $v_{s1}$ and $v_{s2}$, and knowing the measured engineering shear rate $\gammap$, we can compute
the ``true'' shear rate  $\gammapeff$ produced by the lamellar phase and defined consistently with
Eq.~(\ref{e.gammarheo}) as
\begin{equation}
\gammapeff=\frac{R_1^2+
R_2^2}{R_1 (R_1+R_2)}\,\frac{v_0-v_{s1}-\frac{R_1}{R_2}v_{s2}}{e}\,,
\label{e.effectiveshear}
\end{equation}
where $v_0=R_1\Omega$ is computed from $\gammap$ using Eq.~(\ref{e.gammarheo}) (see also the Appendix).
We recall that the measured engineering shear rate $\gammap$
is the sum of the contributions of the two lubricating layers at the walls and of the lamellar phase.
As seen on Fig.~\ref{f.gliss2}(b), the true shear rate fluctuates much more than the engineering shear rate:
$\delta\gammapeff/\gammapeff=5.5$~\% whereas $\delta\gammap/\gammap=\delta v_0/v_0=2.5$~\% in this case.
Moreover, the engineering shear rate and the true shear rate
may be completely out of phase due to the dynamics of slip velocities:
$\gammapeff$ sometimes decreases whereas $\gammap$ increases as well as the slip velocities. 
This is the case for instance around $t=400$~s and $t=1200$~s.
Thus, wall slip and bulk dynamics are decoupled under imposed shear stress,
leading to the observed discrepancy between the
$\gammap$ and $\gammapeff$ signals. Note that, if the experiments were conducted under applied
shear rate ({\it i.e.} at fixed $v_0$), the dynamics of $\gammapeff$ would be automatically
coupled to that of $v_{s1}+v_{s2}$ through Eq.~(\ref{e.effectiveshear}).

\begin{figure}[htbp]
\begin{center}
\scalebox{1}{\includegraphics{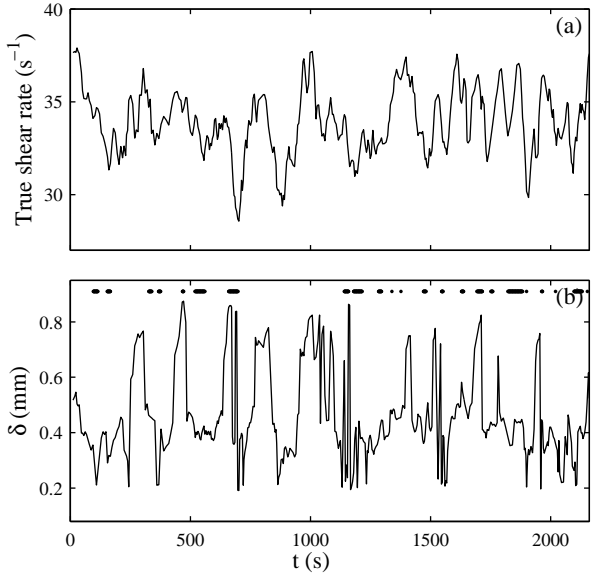}}
\end{center}
\caption{\label{f.bande}(a) True shear rate $\gammapeff(t)$. (b) Position of the interface $\delta(t)$. 
Black lines and dots indicate time intervals when two interfaces are present (see text).}
\end{figure}

\begin{figure}[htbp]
\begin{center}
\scalebox{1}{\includegraphics{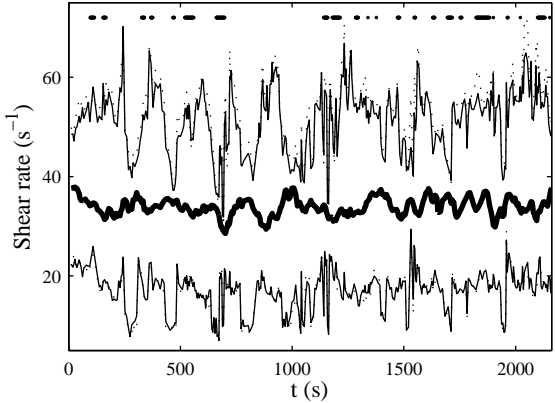}}
\end{center}
\caption{\label{f.cisloc2}True shear rate $\gammapeff(t)$ (thick line) and
local shear rates $\gammap_1(t)$ (thin bottom line) and $\gammap_2(t)$ (thin top line) in the weakly and highly sheared band
computed from Eqs.~(\ref{e.cisloc1}) and (\ref{e.cisloc2}) respectively. The dotted lines correspond to estimations
of $\gammap_1(t)$ and $\gammap_2(t)$ by averaging the local shear rate $\gammap(x,t)$ in the two bands.
Black lines and dots indicate time intervals when the analysis is not strictly valid due to the presence
of more than two shear bands (see text).}
\end{figure}

Figure~\ref{f.bande} presents the position of the main interface as a function of 
time together with the evolution of the true shear rate. 
The position $\delta(t)$ of the main interface is defined as the position
where $\partial\gammap(x,t)/\partial x$ reaches a maximum as a function of $x$. 
The black lines and dots in Fig.~\ref{f.bande}(b) indicate the time intervals when this detection method may not be valid due to
the presence of two interfaces in the gap of the Couette cell.
Two major conclusions may be drawn from these data. 

(i) First, the position of the interface $\delta(t)$ seems rather well correlated
to the true shear rate $\gammapeff(t)$ when a single interface is present in the gap (at least much better correlated than to $\gammap(t)$).
When $\gammapeff(t)$ increases, the band moves towards the stator and tries to fill the entire gap.
When $\gammapeff(t)$ decreases, the band moves back to the rotor.
This is consistent with a dynamical picture of shear banding where fluctuations of the interface position $\delta(t)$ and
of $\gammapeff(t)$ would be linked instantly: an increase of $\gammapeff(t)$ corresponds to an increase of the
proportion of the highly sheared band and thus to an increase of $\delta(t)$ (see also Sec.~\ref{s.discuss}).

(ii) Second, such a correlation between $\gammapeff(t)$ and $\delta(t)$ is sometimes far from perfect,
even when only one interface is present (see $t=1600$--1800~s),
suggesting that the local shear rates in the two bands also fluctuate
in time. Indeed, if the two local shear rates were constant and equal to $\gammap_1$ and $\gammap_2$
in the weakly and highly sheared bands respectively, the continuity of the velocity field would impose that $\delta(t)$ 
and $\gammapeff(t)$ are linearly linked by
\begin{equation}
\gammapeff(t)\simeq\gammap_1\,\frac{e-\delta(t)}{e} + \gammap_2\,\frac{\delta(t)}{e}=\gammap_1+(\gammap_2-\gammap_1)\,\frac{\delta(t)}{e}\,,
\label{e.gammapeffdelta}
\end{equation}
at least in the presence of a single interface.

Figure~\ref{f.cisloc2} presents the evolution of the local shear rates $\gammap_1(t)$ and $\gammap_2(t)$ in the two bands
inferred from $v_{s1}(t)$, $v_{s2}(t)$, $\delta(t)$, and $v(\delta(t),t)$ according to Eqs.~(\ref{e.cisloc1}) and (\ref{e.cisloc2})
(see Appendix).
Another estimation of $\gammap_1(t)$ and $\gammap_2(t)$, obtained by averaging
the local shear rate $\gammap(x,t)$ over $\delta+100~\mu$m$<x<e-100~\mu$m and $100~\mu$m$<x<\delta-100~\mu$m 
respectively, is shown in dotted lines in Fig.~\ref{f.cisloc2}. The two estimations of the local shear rates in the bands
are in good quantitative agreement.
In both bands, the local shear rate strongly fluctuates by $\delta\gammap_1/\gammap_1\simeq 23$~\% and
$\delta\gammap_2/\gammap_2\simeq 12$~\%. Once again, the time scales involved in these fluctuations range from 50 to 500~s.
Such large fluctuations imply that Eq.~(\ref{e.gammapeffdelta}) with constant $\gammap_1$ and $\gammap_2$ does not hold in our experiment.
Finally, note that $\gammap_1(t)$ and $\gammap_2(t)$ are clearly
correlated but seem anti-correlated with $\delta(t)$.
Moreover, the relative fluctuations of the local shear rates
are about twice as large in the weakly sheared band
as in the highly shear band (see Sec.~\ref{s.discuss} for more details).

In this first experiment, we have analyzed the flow dynamics in a regime of small amplitude oscillations. One may also wonder what happens
when large amplitude oscillations are produced such as those observed in
Ref.~\cite{Wunenburger:2001}: what are the similarities and the differences
between those two dynamical regimes? In the following section, we show how to induce easily large
amplitude oscillations and analyze the dynamics involved in such conditions.


\section{Large amplitude oscillations}
\label{s.large}

\subsection{Experimental procedure and rheological measurements}

In this section, we study the regime of the large amplitude oscillations.
This regime is difficult to attain experimentally
by increasing the applied shear stress step by step as in standard procedures:
in most experiments, just after the region of small amplitude oscillations, the system jumps directly to the highly sheared branch and
the regime of large oscillations is missed.
This seems to be related to the very small size of the coexistence region in the orientation diagram.
Indeed, the large oscillations are confined in a metastable region of applied shear stress which is less than 1~Pa wide \cite{Wunenburger:2001,Salmon:2002}.
Moreover, due to filling conditions of the Couette cell that may vary from one experiment to the other,
the exact value of the shear stress that has to be applied to obtain such oscillations is not reproducible.

\begin{figure}[htbp]
\begin{center}
\scalebox{1}{\includegraphics{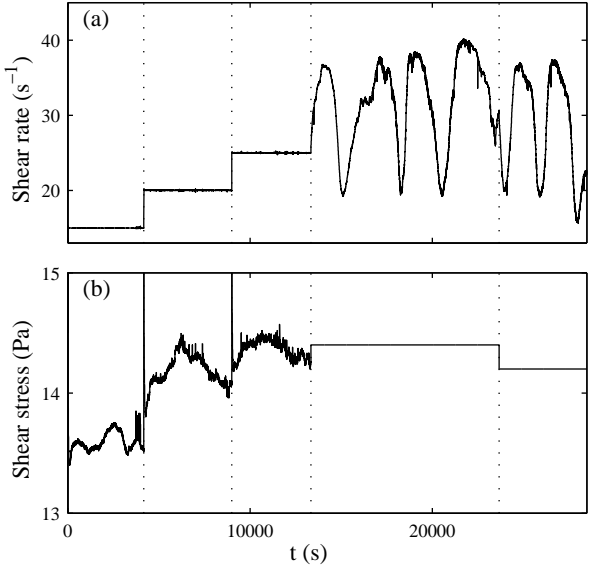}}
\end{center}
\caption{\label{f.rheol}(a) Engineering shear rate $\gammap(t)$. (b) Engineering shear stress $\sigma(t)$.
The dotted lines correspond to the end of the various steps.
The temperature is $T=28~^\circ$C.}
\end{figure}

In order to find this region more easily, we first impose a constant shear stress of 12~Pa for one hour,
followed by a constant shear stress of 13~Pa for another hour at a temperature $T=28~^\circ$C.
This allows us to prepare a stationary state of disordered onions.
We then bring the system into the coexistence domain by applying steps of constant shear rate at $\gammap=15$, 20, and 25~s$^{-1}$ during about 4000~s.
This procedure is easy because the coexistence zone is very large when the applied parameter is the shear rate rather than the shear stress.
We then note the measured shear stress ($\sigma\simeq 14.4$~Pa at $\gammap=25$~s$^{-1}$), switch
the rheometer from the shear-rate-controlled mode to the stress-controlled mode, and apply a constant stress ($\sigma=14.4$~Pa) for about
three hours.
Note that the AR1000N rheometer allows us to program a new step in the rheological procedure without stopping 
the experiment. This allows us to begin the experiment in the coexistence domain at imposed shear stress,
which induces the large oscillations of the viscosity shown in Fig.~\ref{f.rheol}(a): a low viscosity ({\it i.e.} a high shear rate) is observed when the
onions are in the ordered state whereas a high viscosity ({\it i.e.} a small shear rate) is measured when the
onions are in the disordered state. The period of these large oscillations is about 2000~s and their amplitude is
$\delta\eta/\eta=\delta\gammap/\gammap=25$~\%.
Finally, a last step at $\sigma=14.2$~Pa is performed, that shows
that large oscillations are still observed at a slightly lower imposed shear stress.
Figure~\ref{f.rheol} shows the full rheological data sets $\gammap(t)$ and $\sigma(t)$ recorded during such a procedure.

\subsection{Time-averaged velocity profiles}
\label{s.large.average}

\begin{figure}[htbp]
\begin{center}
\scalebox{1}{\includegraphics{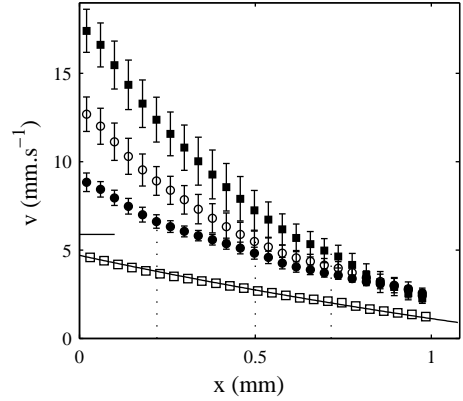}}
\end{center}
\caption{\label{f.shbd}Time-averaged velocity profiles for
$\sigma=12$~Pa ($\square$), $\gammap=15$ ($\bullet$), 20 ($\circ$), and 25~s$^{-1}$ ($\blacksquare$).
The corresponding rheological data are shown in Fig.~\ref{f.rheol} except for $\sigma=12$~Pa.
Error bars correspond to the standard deviation of the local velocities and mainly account for temporal fluctuations of the velocity field.
The solid line corresponds to the homogeneous flow of a power-law fluid (shear-thinning exponent $n=0.2$
and parameter $A=9.3$) with wall slip (see text).
The dotted lines show the average position of the interface between the two shear bands. The mean velocity of the rotor 
is $v_0=5.9$, 14.7, 19.6, and 24.5~mm.s$^{-1}$ respectively. The horizontal solid line indicates $v_0=5.9$~mm.s$^{-1}$.}
\end{figure}

Simultaneously to the rheological measurements, we record the velocity profiles.
Figure~\ref{f.shbd} presents averaged velocity profiles obtained
during pre-shearing at low imposed shear stress ($\sigma=12$ Pa) and
under imposed shear rate.
When $\sigma=12$~Pa, and thus at very low shear rate, the flow is homogeneous and stationary.
Moreover, the velocity profile is curved, revealing that disordered onions are highly shear-thinning.
We also note that the onions slip at both walls.
Indeed, as shown by the solid line in Fig.~\ref{f.shbd},
the velocity profile may be well fitted using a power-law fluid $\sigma=A\gammapeff^n$ with $n=0.2$, as 
long as slip velocities are accounted for.

When the shear rate is applied and increased above $\gammap_1$, the flow becomes inhomogeneous.
A high-shear band nucleates at the rotor and coexists with a low-shear band.
Moreover, the velocity profiles as well as the position of the interface fluctuate,
as shown by the large errors bars on Fig.~\ref{f.shbd} localized near the mean position of the interface.
These fluctuations induce small variations of the measured shear stress ($\delta\sigma/\sigma\simeq 3$~\%)
as seen in Fig.~\ref{f.rheol}(b).
The same behaviour is encountered for all the steps at imposed shear rate in the coexistence region
between disordered and ordered onions \cite{Salmon:2003e}.
Increasing the shear rate in this region only changes the proportion of the two structures.
Moreover, the evolution of the average slip velocities as $\gammap$ is increased is as follows:
$v_{s1}=5.6$, 6.5, and 6.5~mm.s$^{-1}$ 
and $v_{s2}=2.4$, 2.0, and 1.4~mm.s$^{-1}$ for
$\gammap=15$, 20, and 25~s$^{-1}$ respectively.

Finally, once the rheometer is switched to the stress-controlled mode, the fluctuations of the velocity 
field get so large that a global time-average of the velocity profiles becomes meaningless.
In the following section, we provide a detailed analysis of the last step at an imposed shear stress of 14.2~Pa.
Qualitatively, the same behaviour is found when $\sigma=14.4$~Pa is applied.

\subsection{Study of the step at $\sigma=14.2$~Pa}

\subsubsection{Velocity profiles $v(x,t)$}

\begin{figure}[htbp]
\begin{center}
\scalebox{1}{\includegraphics{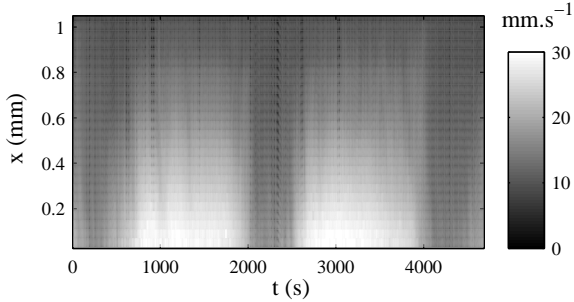}}
\end{center}
\caption{\label{f.sptp}Spatio-temporal diagram of the local velocity $v(x,t)$.
A moving average over three consecutive profiles was used to smooth the experimental data.}
\end{figure}

\begin{figure}[htbp]
\begin{center}
\scalebox{1}{\includegraphics{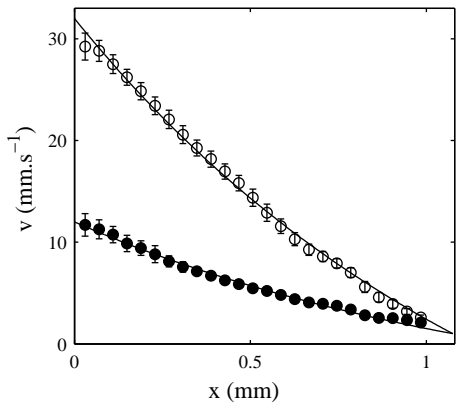}}
\end{center}
\caption{\label{f.moy}Velocity profiles averaged over $t=1122$--1577~s in the
low-viscosity ordered state ($\circ$) and over $t=2172$--2310~s in the high-viscosity disordered state ($\bullet$)
(see the horizontal bars in Fig.~\ref{f.gliss}(b)).
The solid lines correspond to homogeneous flows of power-law fluids (shear-thinning exponent $n=0.1$ in both cases 
and parameters $A=10.15$ and 11.25 respectively) with wall slip (see text).}
\end{figure}

Figure~\ref{f.sptp} presents the spatio-temporal diagram of $v(x,t)$ corresponding
to 760 profiles recorded every 6~s at $\sigma=14.2$~Pa. These measurements reveal that
the velocity profiles oscillate between two homogeneous states: a high and a low viscosity state \cite{Remark:moviesUSV}. 
Figure~\ref{f.moy} shows two velocity profiles averaged in the high and low viscosity states.
The time intervals on which the average was taken are indicated by horizontal bars in Fig.~\ref{f.gliss}(b).
In both cases, the profiles are homogeneous. Moreover, the lamellar phase is highly shear-thinning
since the velocity profiles may be well accounted for by power-law fluids ($\sigma=A\gammapeff^n$) with exponent $n=0.1$.
We can also note that the onions still slip at the two walls of the Couette cell, so that the average slip velocities had
to be included in the fits of Fig.~\ref{f.moy}.

These homogeneous profiles are stable during at least twenty minutes and then suddenly become unstable in a few minutes.
The crossovers between the two states are very fast compared to the period of time spent in the two homogeneous states.
As in Sec.~\ref{s.slip}, we need to quantify the dynamics at the walls before trying to describe precisely the flow field during
the crossovers.

\subsubsection{Wall slip dynamics}

\begin{figure}[htbp]
\begin{center}
\scalebox{1}{\includegraphics{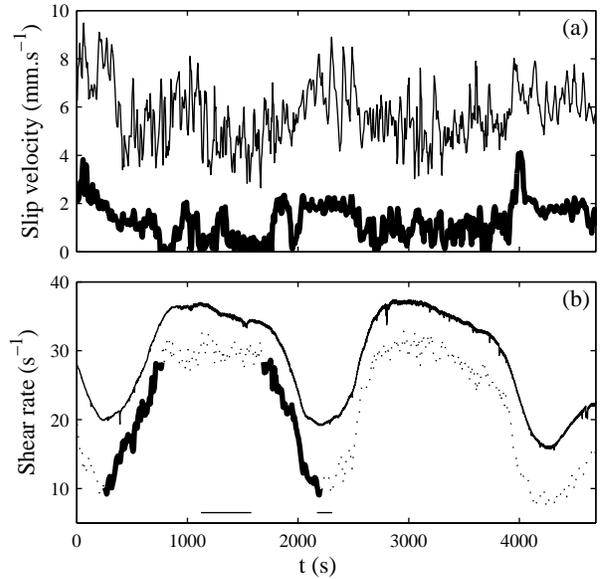}}
\end{center}
\caption{\label{f.gliss}(a) Slip velocities at the rotor $v_{s1}(t)$ (thin line) and at the stator $v_{s2}(t)$ (thick line).
(b) Engineering shear rate $\gammap(t)$ (solid line) and true shear rate $\gammapeff(t)$ (dotted line).
The horizontal bars indicate the time intervals on which the velocity profiles of Fig.~\ref{f.moy} were averaged.
The parts in thick lines refer to the increase and the decrease of the shear rate described
in Figs.~\ref{f.mont} and \ref{f.desc} respectively.}
\end{figure}

Figure~\ref{f.gliss} presents the slip velocities at the rotor and at the stator.
Again, $v_{s1}$ and $v_{s2}$ are significantly different.
Fast fluctuations on time scales shorter than one minute are observed. Slower variations occuring on a time scale of about 1000~s
are superimposed to the fast fluctuations. This long time scale is clearly correlated to the state of the lamellar phase:
both slip velocities are larger (resp. smaller) when the lamellar phase is in the disordered (resp. ordered) state.

Finally, using $v_{s1}(t)$, $v_{s2}(t)$, $\gammap(t)$, and Eq.~(\ref{e.gammapeffdelta}),
we have calculated the true shear rate $\gammapeff(t)$ produced by the lamellar phase
(see Fig.~\ref{f.gliss}(b)). Contrary to what happens in the regime of small amplitude oscillations,
$\gammapeff$ oscillates in phase with $\gammap$ and its variations are very large compared
to the variations of slip velocities.

\subsubsection{Crossover from the weakly sheared state to the highly sheared state}

\begin{figure}[htbp]
\begin{center}
\scalebox{1}{\includegraphics{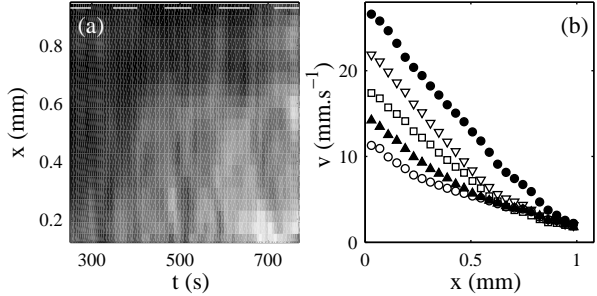}}
\end{center}
\caption{\label{f.mont}Crossover during the increase of the shear rate indicated by a thick line in Fig.~\ref{f.gliss}(b). (a) Local
shear rate $\gammap(x,t)$. White lines indicate the time intervals when the velocity profiles presented in (b) were averaged.
A linear gray scale is used to code the values of $\gammap$: black corresponds to $\gammap=5$~s$^{-1}$ and white corresponds 
to  $\gammap=45$~s$^{-1}$. (b) Velocity profiles averaged over $t=251$--298~s ($\circ$),
348--403~s ($\blacktriangle$), 465--526~s ($\square$),
588--658~s ($\triangledown$), and 714--770~s ($\bullet$).}
\end{figure}

Figure~\ref{f.mont} shows that the system goes from the weakly sheared state to
the highly sheared state via an inhomogeneous flow. A high-shear band is nucleated at the 
rotor and progressively fills the entire gap.
Indeed, when averaging the velocity profiles on about 60~s (as shown by the white lines in Fig.~\ref{f.mont}(a)),
one can see that the local shear rates in the low-shear and high-shear bands first remain constant as the interface between the two
bands expands towards the stator. This occurs for the three first profiles in Fig.~\ref{f.mont}(b), which are
consistent with ``frozen" snapshots of a classical shear-banded flow at different increasing imposed shear rates.

However, the nucleation and displacement of the band does not occur smoothly during all the crossover.
At $t\simeq 500$~s, the interface stops and the viscosity in the highly sheared band decreases
whereas the viscosity in the weakly sheared band remains constant.
Then, the interface moves towards the stator again and the viscosity in the high-shear band does not change anymore.
At the end of this process, we are left with a homogeneous velocity profile.

Even if such a process is not exactly reproduced from one crossover to the other, the general features of the transition
from the weakly sheared state to the highly sheared state remain the same for all the investigated crossovers: steady
growth of the band at fixed $\gammap_1$ and $\gammap_2$ seems to alternate with periods where the interface remains fixed while
the local viscosities vary.

\subsubsection{Crossover from the highly sheared state to the weakly sheared state}
\begin{figure}[htbp]
\begin{center}
\scalebox{1}{\includegraphics{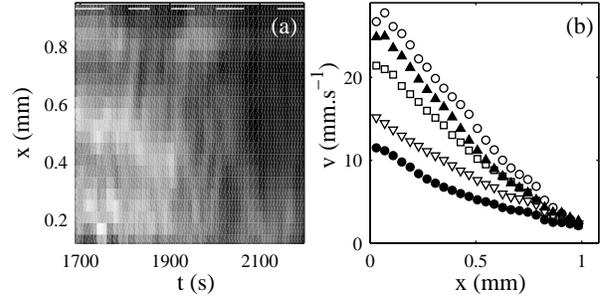}}
\end{center}
\caption{\label{f.desc}Crossover during the decrease of the shear rate indicated by a thick line in Fig.~\ref{f.gliss}(b). (a) Local
shear rate $\gammap(x,t)$. White lines indicate the time intervals when the velocity profiles presented in (b) were averaged.
A linear gray scale is used to code the values of $\gammap$: black corresponds to $\gammap=5$~s$^{-1}$ and white corresponds 
to  $\gammap=45$~s$^{-1}$. (b) Velocity profiles averaged over $t=1686$--1754~s ($\circ$),
1807--1855~s ($\blacktriangle$), 1903--1955~s ($\square$),
2003--2065~s ($\triangledown$), and 2139--2198~s ($\bullet$).}
\end{figure}
However, the above picture does not hold anymore during the crossover from the low viscosity state to the high viscosity state.
In this case, as seen on Fig.~\ref{f.desc}, the flow remains homogeneous
across the Couette cell as the shear rate decreases. Thus, onions become disordered simultaneously everywhere in the gap.
The viscosity progressively increases without any significant inhomogeneities and the true shear rate decreases accordingly.

Once again, even if inhomogeneous flows are sometimes observed transiently on individual profiles during the return
to the disordered state, this general picture holds for the various investigated crossovers
from the highly sheared state to the weakly sheared state. 


\section{Discussion and model}
\label {s.discuss}

\subsection{Summary}

Let us now summarize the results obtained in the present work.
Under applied shear stress and inside the coexistence region of the layering transition,
the measured engineering shear rate oscillates and the flow is inhomogeneous.
Two bands (or even more) of different viscosities
coexist in the gap of the Couette cell and their relative amount varies as a function of time. 
Wall slip is also very important. The recorded signal $\gammap(t)$
is the sum of three separated dynamics: the dynamics of the two slip velocities
and the dynamics of the lamellar phase.

In the bulk, the true shear rate $\gammapeff(t)$ varies mainly because the interface between the two
differently sheared bands oscillates. To a lower extent, the variations of $\gammapeff(t)$ are also linked to
the fluctuations of the local viscosities ({\it i.e.} of the local shear rates) in the two bands. In some particular cases,
a highly sheared band is nucleated close to the stator. This secondary high-shear band is unstable
and either dies at the stator or moves rapidly towards the rotor where it coalesces with the initial highly sheared region.
Most of the above features are shared by both the small amplitude oscillations and the large amplitude oscillations.
The main difference between large oscillations and small oscillations lies
in the fact that large oscillations require the displacement of the band across the entire Couette cell whereas
during small oscillations, the flow is always inhomogeneous and the interface between bands
fluctuates only in one part of the cell. Moreover, in the regime of large oscillations, a dissymmetry is observed
between the crossovers from one homogeneous state to the other: the apparition of the ordered state seems to follow a
nucleation and growth scenario whereas its destruction occurs homogeneously across the gap.

\subsection{Discussion}

Let us discuss our results in the existing experimental and theoretical framework.
First note that these results are in agreement with our former analysis of the rheological data using dynamical system theory \cite{Salmon:2002}.
In this work, we were unable to prove that the dynamics of $\gammap(t)$ corresponds to a three-dimensional deterministic
chaotic system even though it has several distinctive features resembling those of a low-dimensional chaotic system. 
We proposed that spatial degrees of freedom were responsible for this failure.
The present experimental results clearly show that spatial degrees of freedom are indeed involved: the recorded dynamics involves (at least)
three different dynamics that seem uncoupled to each other. However, it remains to be explained why so many different regimes are
observed, some of them strikingly close to temporal chaos, and others looking much more turbulent.

Second, let us see whether our experimental data are consistent with
the main theoretical approaches dealing with complex fluids under shear flow.
In the last decade, various microscopic or phenomenological
approaches as well as numerical simulations have been used to describe
inhomogeneous flows
of sheared complex fluids
\cite{Olmsted:1992a,Goveas:2001,Picard:2002,Varnik:2003}.
In those works, shear banding is captured by invoking
multiple flow branches in the constitutive relation $\sigma$ vs.
$\gammap$, corresponding to different microstructures.
Under shear, the system separates into a steady state composed of
shear bands, each lying on its own flow branch.
The presence of spatial derivatives in the dynamical equations of motion
selects a unique stress $\sigma^\star$
at which the interface between two shear bands may be stable
\cite{Goveas:2001,Olmsted:1999b,Dhont:1999,Lu:2000}.
This explains the experimental degeneracy in the shear stress at which
coexistence occurs and assumes
that the interface between the two bands is stable only at
$\sigma^\star$. In the Couette geometry, the stress
inhomogeneity implies that the highly sheared band nucleates at
the rotor and is separated from the weakly sheared region by a single
interface. It also leads to the following relationship between the
shear stress at the rotor $\sigma_1$
and the interface position $\delta$ inside the gap \cite{Salmon:2003d}:
\begin{equation}
\delta=\left(\sqrt{\frac{\sigma_1}{\sigma^\star}}-1\right)\,R_1\,.
\label{eq.sigmastar}
\end{equation}

In a previous work, we have shown that such a shear banding
scenario holds for the layering transition in our
lamellar phase \cite{Salmon:2003d}.
It describes quantitatively the time-averaged profiles under both applied shear rate
and shear stress (at least in the region of the small amplitude oscillations).
However, this oversimplified picture does not include the possibilty of oscillations of the rheological parameters
and of sustained oscillations of the position of the interface between the shear bands.
We have proposed that these oscillations may be captured by assuming that
the selected stress $\sigma^\star$ is a fluctuating parameter in our system \cite{Salmon:2003e}.
Under the assumption of a fluctuating $\sigma^\star$, we may describe at least qualitatively the dynamics.
Indeed, the same small fluctuations of $\sigma^\star$ will produce small variations of the
measured shear stress in the case of applied shear rate but huge variations
of the measured shear rate in the case of applied shear stress.
Thus, a microscopic equation was still missing to describe the evolution of $\sigma^\star$.

In the following, we first focus on the nonlocal approach of shear-banded
flows developed in Refs.~\cite{Olmsted:1999b,Dhont:1999}.
We then propose a simple set of equations based on such an approach but including the
possibility of temporal dynamics.
This crude dynamical model may provide an explanation for variations
of $\sigma^\star$ and for the observed dynamics.

\subsection{The nonlocal approach of Refs.~\cite{Olmsted:1999b,Dhont:1999}}

In Refs.~\cite{Olmsted:1999b,Dhont:1999}, it is assumed
that the stress in the system is locally
the sum of two terms, one related to the shear rate and the other one
proportional to the second derivative of the shear rate:
\begin{equation}
\sigma=g(\gammap)-\kappa\, \frac{\partial^2\gammap}{\partial x^2}\,.
\label{e.sigmadhontzero}
\end{equation}
The first term corresponds to the usual constitutive equation for a
homogeneous flow. The friction $g(\gammap)$ between two adjacent
sliding layers depends only on their relative velocity {\it i.e.} on
the local shear rate. However, when large spatial variations of the local
shear rate occur, a second term must be added that involves
$\partial^2\gammap/\partial x^2$.
In this latter case, the microstructure of the fluid
varies on a length scale of the order of the range of interactions
between the mesoscopic entities of the system ({\it i.e.}
the microstructure varies on the scale of a
hydrodynamic layer). To take into account these inhomogenities of the
microstructure,
Dhont has introduced the ``shear curvature viscosity'' $\kappa$
(see Eq.~(\ref{e.sigmadhontzero})) \cite{Dhont:1999}.

Like the standard shear viscosity, such a shear curvature viscosity depends
upon the shear rate. In particular, since the microstructure generally reaches
a homogeneous state at high shear rate,
$\kappa$ has to decrease to zero when $\gammap$ goes to infinity. For
example, for a fluid of rodlike molecules, shear progressively
aligns the rods in the flow direction. At high shear rate, the rods
are fully aligned and the microstructure cannot evolve further.
Thus, the fluid becomes more homogeneous and the shear
curvature viscosity has to vanish.
To describe this behaviour, the authors propose that
\begin{equation}
\kappa=\frac{D}{1+\alpha\,\gammap^2}\,,
\label{e.alphadhont}
\end{equation}
where $D$ is analogous to a diffusion coefficient
and $\alpha$ is a phenomenological coefficient \cite{Olmsted:1999b,Dhont:1999}.

Let us now analyze the consequences of such a behaviour
in the presence of a shear-induced transition.
In this case, the first term $g(\gammap)$
is a multivalued function of the shear stress:
for a small range of applied shear stresses, two branches of different
structures exist.
In a plane geometry, Eqs.~(\ref{e.sigmadhontzero}) and (\ref{e.alphadhont}) select a
unique stress $\sigma^\star$
for which the two different structures coexist.
If the applied shear stress is smaller than this selected
stress $\sigma^\star$, the system remains on the low-shear branch.
Above $\sigma^\star$, the structure changes and the system jumps
on the highly sheared branch. The value of $\sigma^\star$ is found by
solving the following system of equations \cite{Olmsted:1999b,Dhont:1999}:
\begin{eqnarray}
\int^{\gammap_2}_{\gammap_1}{\left( \sigma^\star-
g(\gammap)\right)\left(1+\alpha\,\gammap^2\right)
\hbox{\rm d}\gammap} =0\,,\label{eq.dhont1}\\
\sigma^\star=g(\gammap_1)=g(\gammap_2)\,,
\label{eq.dhont2}
\end{eqnarray}
where the unknowns $\gammap_1$ and $\gammap_2$ are the
local shear rates in the low-shear and
high-shear bands respectively.

\subsection{Toy model}

The nonlocal approach described above allows a good description of
shear-banded flows but is intrinsically {\it stationary} and
unable to reproduce any of the dynamics reported in the experiments.
Recently, various phenomenological models have been proposed to
mimic the chaotic-like spatio-temporal behaviours of sheared complex
fluids, mostly in the context of wormlike micelles and colloidal pastes
\cite{Cates:2002,Fielding:2003pp,Aradian:2003pp}, as well as
liquid crystalline polymers \cite{Grosso:2001,Grosso:2003,Chakrabarti:2003pp}.

Here, we adapt the model of Refs.~\cite{Olmsted:1999b,Dhont:1999} by taking into account
the particularity of our system, namely an
assembly of soft multilamellar vesicles (onions) whose interlayer distance
may vary under shear.
Indeed, using neutron scattering, it has been measured that the
smectic period of the ordered onion texture decreases
as the shear rate is increased whereas it remains constant in the disordered
texture \cite{Leng:2001}.
It was assumed that some water is expelled from the onions and lies between the different
planes of ordered onions to lubricate the layers.
The time scale involved in such a release of water
depends on the temperature, due to the presence of thermally activated defects in
the lamellar phase. It ranges from ten minutes to an hour.
Those water layers dramatically affect the viscosity of the sample.

In the framework of Refs.~\cite{Olmsted:1999b,Dhont:1999}, such water layers should
also strongly influence the coefficient $\alpha$ in the shear curvature viscosity
(see Eq.~(\ref{e.alphadhont})). A scenario leading to sustained oscillatory
behaviour then takes shape.
Indeed, a decrease of $\alpha$, linked to the slow release of water, may
induce an increase in $\sigma^\star$.
Consequently, an increase of $\sigma^\star$ may lead to the destruction of
the shear-induced phase, decrease the amount of water between the onions,
and thus increase $\alpha$ (see below).
The competition between those two processes may 
induce oscillations of $\sigma^\star$ and thus lead to oscillations of the
measured shear rate under applied shear stress.

More precisely, in order to mimic these two antagonist effects, we introduce
$h(t)$ the total amount of water lubricating the planes of ordered onions.
For a stationary state in the coexistence region, we assume that
$h$ is directly proportional to the amount of ordered onions $\epsilon$.
When the flow is composed of a single interface separating a highly-sheared
ordered region from a weakly-sheared disordered region, one simply has
$\epsilon=\delta/e$, where $\delta$ is the position of the interface
inside the gap of width $e$.
We then write two coupled
phenomenological equations obeying the following rules.

(i) $h$
must relax to $\heq=0$ in the absence of the shear-induced structure, to some
maximum value $\heq=h_0$ when the fluid is fully ordered, and to the
intermediate value $\heq=\epsilon h_0$ in the coexistence region.

(ii) $\alpha$ must decrease when $h$ increases. Indeed,
when water is released, {\it i.e.} when $h$
increases, the microstructure of the ordered phase differs more sharply
from that of the disordered phase. This should induce an increase of
the shear curvature viscosity $\kappa$, which in our
model corresponds to a decrease of $\alpha$. Thus, we assume that, at equilibrium,
$\alpha$ relaxes to $\alpha_0-\beta\heq$,
where $\alpha_0$ and $\beta$ are positive parameters.
Assuming exponential relaxations with characteristic times $\tau_1$
and $\tau_2$, we propose to describe these two processes by
\begin{eqnarray}
\frac{\hbox{\rm d}h}{\hbox{\rm d}t}&=&\frac{1}{\tau_1}\left(\epsilon(t) h_0 -h(t)\right)\,,
\label{eq.dyn1}\\
\frac{\hbox{\rm d}\alpha}{\hbox{\rm d}t}&=&\frac{1}{\tau_2}\left(\alpha_0-\beta h(t)
-\alpha(t)\right)\,.
\label{eq.dyn2}
\end{eqnarray}

Moreover, if Eq.~(\ref{eq.sigmastar}) holds at all times, then $\epsilon(t)$ is given by
\begin{equation}
\epsilon(t)=\frac{\delta(t)}{e}=\left(\sqrt{\frac{\sigma_1}{\sigma^\star(t)}}-1\right)\,
\frac{R_1}{e}\,,
\label{e.sigmastarbis}
\end{equation}
where the stress at the rotor $\sigma_1$ is linked to the applied shear stress
by Eq.~(\ref{e.sigmawall}). At each time step, $\sigma^\star(t)$ is
found by solving Eqs.~(\ref{eq.dhont1})--(\ref{eq.dhont2})
using the current value $\alpha(t)$.
The instantaneous relation (\ref{e.sigmastarbis})
between $\epsilon(t)$ and $\sigma^\star(t)$ simply
means that the time scales involved in the selection of $\sigma^\star$ and
in the transport of the shear stress are short compared to the
swelling and deswelling processes of the onion phase.
A more realistic approach would involve a third dynamical equation
coupling $\delta$ and $\sigma^\star$ and/or memory effects.
This could lead to more complex dynamics than the simple oscillatory state found
here, possibly to chaotic dynamics \cite{Cates:2002}. However, such an
extension of the model is beyond the scope of the present article.

In order to solve Eqs.~(\ref{eq.dhont1})--(\ref{e.sigmastarbis}),
a modelization of the rheological curve $g(\gammap)$ describing the homogeneous
part of the flow is required. For the sake of simplicity,
we assume that the two branches are Newtonian and we define $g(\gammap)$ as
\begin{eqnarray}
\ g(\gammap) &=& \eta_1 \gammap\,\,\,\hbox{\rm if}\,\,\,\gammap\leq \gammap_c\,,
\label{eq.ecoul1}\\
g(\gammap) &=& \eta_2 \gammap\,\,\,\hbox{\rm if}\,\,\,\gammap\geq \gammap_c \,,
\label{eq.ecoul2}
\end{eqnarray}
where $\gammap_c$ is some critical shear rate
and $\eta_1$ and $\eta_2$ are the viscosities of the
disordered and ordered states respectively ($\eta_1>\eta_2$). Thus,
for $\sigma\in [\eta_2\gammap_c;\eta_1\gammap_c]$,
two different shear rates are accessible to the system. Eqs. (\ref{eq.ecoul1}) and
(\ref{eq.ecoul2}) provide a coarse approximation
of the experimental flow curve but
allow us to solve analytically Eqs.~(\ref{eq.dhont1})--(\ref{eq.dhont2})
and to implement easily
the numerical resolution of Eqs.~(\ref{eq.dyn1})--(\ref{e.sigmastarbis}).

\begin{figure}[htbp]
\begin{center}
\scalebox{1}{\includegraphics{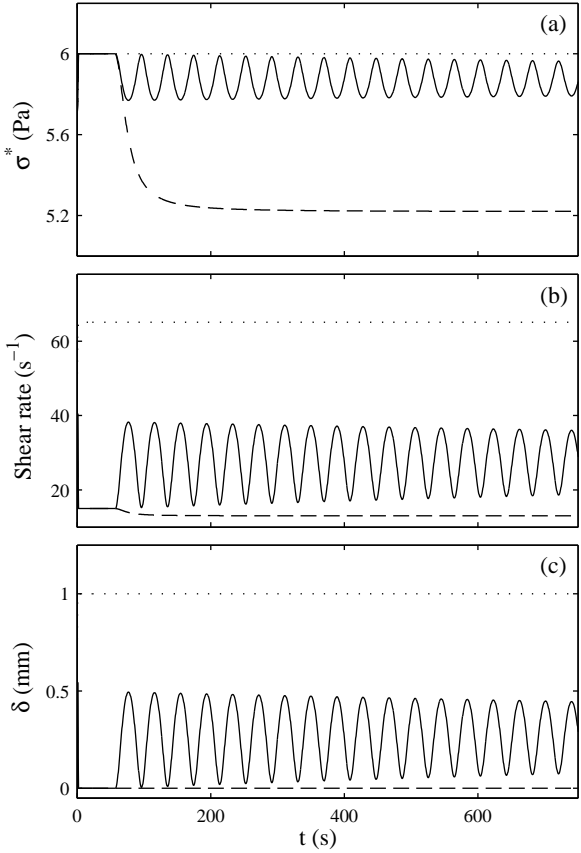}}
\end{center}
\caption{\label{f.model}Solutions of Eqs.~(\ref{eq.dhont1})--(\ref{e.gammapeffdeltabis})
for $\eta_1=0.4$~Pa.s, $\eta_2=0.1$~Pa.s, $\gammap_c=30$~s$^{-1}$,
$\tau_1=\tau_2=200$~s, $h_0=30$, $\alpha_0=0.2$, and $\beta=0.024$.
(a) Selected shear stress $\sigma^\star(t)$. (b) Global shear rate
$\gammap(t)$. (c) Position of the interface $\delta(t)$.
The imposed shear stress corresponds to a shear stress at the rotor
$\sigma_1=5.2$ (dashed line), 6.0 (solid line), and 6.5~Pa (dotted line).
The dimensions of the Couette cell are $R_1=24$~mm and $R_2=25$~mm.}
\end{figure}

Finally, the local shear rates $\gammap_1(t)$
and $\gammap_2(t)$ are recovered using Eq.~(\ref{eq.dhont2})
and the global shear rate $\gammap(t)$ is computed by
\begin{equation}
\gammapeff(t)=\gammap_1(t)+(\gammap_2(t)-\gammap_1(t))\,\frac{\delta(t)}{e}\,.
\label{e.gammapeffdeltabis}
\end{equation}
Note that, in principle, Eqs.~(\ref{e.sigmadhontzero})--(\ref{eq.dhont2}) and
(\ref{e.gammapeffdeltabis}) only hold in a plane geometry,
whereas Eq.~(\ref{e.sigmastarbis}) results from the stress inhomogeneity
inherent to the Couette geometry. However, as shown in the Appendix, the aspect
ratio $e/R_1$ of our Couette cell is small enough
to use Eq.~(\ref{e.gammapeffdelta}) (or equivalently
Eq.~(\ref{e.gammapeffdeltabis})), and more generally, to assume that
Eqs.~(\ref{e.sigmadhontzero})--(\ref{eq.dhont2}) are not significantly altered by
curvature effects.

Figure~\ref{f.model} presents the temporal signals obtained using the following set
of parameters: $\eta_1=0.4$~Pa.s, $\eta_2=0.1$~Pa.s, $\gammap_c=30$~s$^{-1}$,
$\tau_1=\tau_2=200$~s, $h_0=30$, $\alpha_0=0.2$, and $\beta=0.024$. Time series
corresponding to $\sigma^\star(t)$, $\gammap(t)$, and $\delta(t)$ are shown for
various imposed shear stresses corresponding to $\sigma_1=5.2$, 6.0, and 6.5~Pa.
For a small value of the applied shear stress ($\sigma_1=5.2$~Pa, dashed line), the
system relaxes towards a homogeneous stationary state.
The system remains in the low-shear state since the shear rate is simply
$\gammap\simeq\sigma_1/\eta_1\simeq 13$~s$^{-1}$ and $\delta=0$.
For large applied shear stresses ($\sigma_1=6.5$~Pa, dotted line), another homogeneous state is
obtained but the viscosity of the system is now $\eta_2$ since 
$\gammap\simeq\sigma_1/\eta_2\simeq 65$~s$^{-1}$. The fluid is then fully
ordered ($\delta\simeq e$).

\begin{figure}[htbp]
\begin{center}
\scalebox{1}{\includegraphics{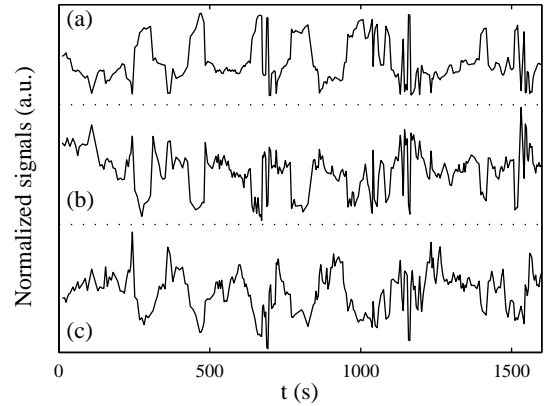}}
\end{center}
\caption{\label{f.compar}Normalized versions of the temporal signals of Figs.~\ref{f.bande} and \ref{f.cisloc2} (see text).
(a) Position of the interface $\delta(t)$. (b) Local shear rate in the weakly sheared band $\gammap_1(t)$.
(c) Local shear rate in the highly sheared band $\gammap_2(t)$.}
\end{figure}

For intermediate values of the applied shear stress ($\sigma_1=6.0$~Pa, solid line),
the two structures coexist and our simple model predicts weakly damped
oscillations. The period of the oscillations depends on the values of
$\tau_1$, $\tau_2$, $\beta$, and
$\partial\sigma^\star/\partial\alpha$.
The interface position $\delta(t)$ oscillates in phase
with $\gammap(t)$ and in phase opposition with $\sigma^\star(t)$ as well as
$\gammap_1(t)=\sigma^\star(t)/\eta_1$ and $\gammap_2(t)=\sigma^\star(t)/\eta_2$.
Experimentally, the same behaviour has already been briefly mentioned in Sec.~\ref{s.slip}.
In Fig.~\ref{f.compar}, we plotted the experimental signals $\delta(t)$, $\gammap_1(t)$, and
$\gammap_2(t)$ normalized by their standard deviations and arbitrarily shifted vertically for
better comparison. It is clearly seen that $\delta(t)$ is indeed anti-correlated with $\gammap_1(t)$ and
$\gammap_2(t)$, which provides a nice confirmation for underlying temporal
variations of $\sigma^\star$.

Note that, in the model, $\sigma^\star(t)$ fluctuates
by about 1.2~\% whereas large oscillations of $\gammap(t)$ are found ($\delta\gammap/\gammap\simeq 25$~\%),
which is consistent with the experimental data. However,
due to the Newtonian assumptions of Eqs.~(\ref{eq.ecoul1}) and (\ref{eq.ecoul2}), the local
shear rates $\gammap_1(t)$ and $\gammap_2(t)$ fluctuate by the same amount as $\sigma^\star(t)$
($\simeq 1.2$~\%), which contradicts the experimental results of Sec.~\ref{s.slip} that show much larger fluctuations.
However, if nonlinear relationships $g(\gammap)=A_i\gammap^{n_i}$
were considered, with $i=1$ (resp. $i=2$) for the low-shear branch (resp. high-shear branch),
Eq.~(\ref{eq.dhont2}) would yield $\delta\sigma^\star/\sigma^\star=n_i\delta\gammap_i/\gammap_i$.
This would lead to fluctuations of $\gammap_i(t)$ enhanced by a factor $1/n_i$ as
compared to those of $\sigma^\star(t)$ and, if $n_1\neq n_2$, to some dissymmetry between the fluctuations
of $\gammap_1(t)$ and $\gammap_2(t)$ as seen in the experiments.
Finally, note that the present model is also tractable under applied shear rate.
Such an analysis is left for future work.


\section{Conclusions}

In this paper, we have shown that our experimental setup based on ultrasonic velocimetry
allows us to follow the dynamics of the inhomogeneous flow of a complex fluid.
We have defined precise rheological procedures in order to produce oscillations of either small or large amplitudes.
Our experimental data unveil two major topics:

(i) The dynamics observed on engineering rheological quantities is the sum of three {\it a priori} uncoupled systems:
the two lubricating layers at the walls and the lamellar phase. This point is very important since it confirms the conclusions
reached after an independent data analysis using dynamical system theory \cite{Salmon:2002}.
In the present study, we have confirmed that the recorded signals cannot result from
a low-dimensional deterministic chaotic system since (at least) three spatial systems are involved.

(ii) The bulk dynamics is mainly due to the displacement
of the interface between two differently sheared bands.
Inspired by the ideas developed in Refs.~\cite{Olmsted:1999b,Dhont:1999},
we proposed that such a motion may be linked to the slow
evolution of the selected stress $\sigma^\star$.
Assuming that the shear curvature viscosity depends upon the amount
of water released by the ordered onions,
we have shown that $\sigma^\star$
could oscillate. These oscillations of $\sigma^\star$
lead to oscillations of the interface position $\delta(t)$ that are in phase opposition with the local shear rates
$\gammap_1(t)$ and $\gammap_2(t)$.
The above features of the model are nicely confirmed by
the experimental signals, even though their dynamics are more complex.

Note that the same kind of arguments may be invoked to
describe the dynamics of the lubricating layers. More elaborate versions
of the equations proposed here may
induce complex behaviours closer to those recorded experimentally.

Future experiments will focus on the measurement of the smectic period under imposed shear stress in the coexistence
region, in order to reach a fully microscopic description of the dynamics. Moreover, using cell walls of controlled
roughness may help us to better understand the origin and the influence of wall slip dynamics.


\begin{acknowledgement}
The authors wish to thank the ``Cellule Instrumentation'' at CRPP
for designing and building the mechanical parts of the experimental setup.
We are very grateful to A. Aradian, L. B{\'e}cu, C. Gay, D. Roux, and A.-S. Wunenburger for fruitful discussions.
\end{acknowledgement}

\section*{Appendix: Determination of the true shear rate $\gammapeff$ and of the local shear rates $\gammap_1$ and $\gammap_2$}
\label{s.appendix}

Let us consider the case of a Newtonian fluid of viscosity $\eta$ sheared in the Couette geometry.
Moreover, let us assume that the fluid slips at both walls so that $v(x=0)=v_1$ and $v(x=e)=v_2$.
Using the local rheological behaviour $\sigma(x)=\eta\gammap(x)$ and
\begin{equation}
\sigma(x) = \frac{\Gamma}{2\pi H (R_1 + x)^2}\, ,
\label{e.sigmax}
\end{equation}
where $\Gamma$ is the torque applied on the rotor axis and $H$ the height of the Couette cell, together with
\begin{equation}
\gammap(x,t)=-(R_1+x)\,\frac{\partial}{\partial x}\,\left( \frac{v(x,t)}{R_1+x}\right)\, ,
\label{e.localshearrate}
\end{equation}
one easily gets the following velocity profile:
\begin{equation}
v(x)=v_2\frac{R_1+x}{R_2}+(R_1+x)\frac{\Gamma}{4\pi\eta H R_2^2}\left[\left(\frac{R_2}{R_1+x}\right)^2 -1 \right]\,.
\label{e.newtprofile}
\end{equation}
Taking $x=0$ in Eq.~(\ref{e.newtprofile}) yields
\begin{equation}
v_1=\frac{R_1}{R_2}\, v_2+\frac{\Gamma}{4\pi\eta H R_1 R_2^2}\,(R_2^2 - R_1^2)\,.
\label{e.v1}
\end{equation}
Finally, with $e=R_2-R_1$, substituting $\sigma$ defined by Eq.~(\ref{e.sigmarheo}) into Eq.~(\ref{e.v1}) leads to
\begin{equation}
\frac{\sigma}{\eta}\,\hat {=}\,\gammapeff=\frac{R_1^2+R_2^2}{R_1 (R_1+R_2)}\,\,\,\frac{v_1-\frac{R_1}{R_2}v_{2}}{e}\,,
\label{e.effectiveshearbis}
\end{equation}
which is exactly Eq.~(\ref{e.effectiveshear}) where
$v_{s1}=v_0-v_1$ and $v_{s2}=v_2$.

In the presence of a two-banded flow, we may now define the local shear rate $\gammap_1$ in the weakly sheared band
from the position $\delta$ of the interface, the velocity $v(\delta)$ measured at $x=\delta$, and the slip velocity $v_{s2}$ at the stator as
\begin{equation}
\gammap_1=\frac{(R_1+\delta)^2+R_2^2}{(R_1+\delta) (R_1+R_2+\delta)}\,\,\,\frac{v(\delta)-\frac{R_1+\delta}{R_2}v_{s2}}{e-\delta}\,,
\label{e.cisloc1}
\end{equation}
Equation (\ref{e.cisloc1}) was simply obtained by taking $R_1\rightarrow R_1+\delta$, $e\rightarrow e-\delta$, and $v_1\rightarrow v(\delta)$
in Eq.~(\ref{e.effectiveshearbis}). Similarly, we define the shear rate $\gammap_2$ in the highly sheared band as
\begin{equation}
\gammap_2=\frac{R_1^2+(R_1+\delta)^2}{R_1 (2 R_1 +\delta)}\,\,\,\frac{v_1-\frac{R_1}{R_1+\delta}v(\delta)}{\delta}\,.
\label{e.cisloc2}
\end{equation}
Combining Eqs.~(\ref{e.effectiveshearbis}), (\ref{e.cisloc1}), and (\ref{e.cisloc2}), it is easy to see that Eq.~(\ref{e.gammapeffdelta})
is not a strict equality but rather results from the small-gap approximation $e/R_1\ll 1$. Actually, using the dimensions of our Couette cell,
we checked that the estimation of $\gammapeff$ given by Eq.~(\ref{e.gammapeffdelta}) differs from that of Eq.~(\ref{e.effectiveshearbis})
by at most 4~\%, which is of the same order of magnitude as the experimental uncertainty on
the velocity measurements.

Finally, let us mention that our fluid is far from being Newtonian since shear-thinning exponents $n=0.1$--$0.2$ are reported
in Sec.~\ref{s.large}. Repeating the above analysis with the local rheological behaviour $\sigma(x)=A\gammap(x)^n$ leads to
\begin{equation}
\gammapeff=\frac{2^{1-1/n}}{nR_1}\frac{(R_1^2+R_2^2)^{1/n}}{R_2^{2/n}-R_1^{2/n}}\,\left(v_1-\frac{R_1}{R_2}v_{2}\right)\,,
\label{e.effectiveshearter}
\end{equation}
instead of Eq.~(\ref{e.effectiveshearbis}).
However, for our Couette cell, the difference between the estimation of Eq.~(\ref{e.effectiveshearbis}) and that of Eq.~(\ref{e.effectiveshearter})
is only 3~\% in the worst case $n=0.1$, which allows us to use Eqs.~(\ref{e.effectiveshearbis})--(\ref{e.cisloc2}) to define $\gammapeff$,
$\gammap_1$, and $\gammap_2$, even if the fluid is not Newtonian.


\begin{thebibliography}{99}
\bibitem{Remark:fluctu}
Throughout this paper, the notation $\delta x/x$ is used to denote the relative
  temporal fluctuations of $x(t)$ defined as $\delta
  x/x=\sqrt{<x^2(t)>-<x(t)>^2}/<x(t)>$, where $<\dots>$ stands for time
  average.
\end{thebibliography}

\end{document}